\documentclass[preprint,12pt]{elsarticle}

\usepackage{amssymb}
\usepackage{amsthm}

\usepackage{enumitem}

\journal{Computer Physics Communication}

\begin{document}

\begin{frontmatter}




\title{{\bf Huygens-Fresnel wavefront tracing in non-uniform media}}


\author{F.A.~Volpe}
\ead{fvolpe@columbia.edu}
\author{P.-D.~L\'{e}tourneau}
\author{A.~Zhao}
\address{Dept of Applied Physics and Applied Mathematics, 
Columbia University, New York, NY}

\begin{abstract}
We present preliminary results on a novel numerical method describing
wave propagation in slowly non-uniform media. Following 
Huygens-Fresnel’s principle, we model the wavefront as an
array of point sources that emit wavelets, which interfere. We  then 
identify a set of new points where the electric field has equal phase. In fact, 
without losing generality, we find zeros of the electric field, by means 
of the bisection method. 
This obviously corresponds to a specific phase-advance, but
is easily generalized, e.g.~by phase-shifting all sources.  
The points found form the new wavefront.  
One of the advantages of the method is that it includes diffraction. 
Two examples provided are diffraction around an obstacle
and the finite waist of a focused Gaussian beam. Refraction is also 
successfully modeled, both in slowly-varying media as well as in 
the presence of discontinuities. The calculations were
performed in two dimensions, but can be easily extended to three
dimensions. We also discuss the extension to anisotropic, birefringent, 
absorbing media.
\end{abstract}


\end{frontmatter}


\section{Introduction} \label{}
Huygens stated in 1690 that ``each element of a wavefront may be regarded 
as the centre of a secondary disturbance which gives rise to spherical 
wavelets''. He added that ``the position of the wavefront at any later time is 
the envelope of all such wavelets'' \cite{Huygens,BornWolf}. This result is 
referred to as Huygens' theorem or construction. 
Indeed we can use it to construct subsequent wavefronts, if we ascertain 
some subtleties in what ``envelope'' means in this context. Such 
subtleties were eventually addressed by Fresnel 
\cite{Fresnel1,Fresnel2,Baker}, as discussed below. 
Also, the construction should be regarded as a 
mathematical abstraction that correctly reproduces the physics 
without necessarily being physically rigorous.  
Treating the wavefront elements as actual centers of secondary 
disturbances might be appropriate in some physics areas, but 
should not be taken literally in optics:
obviously ``light does not emit light; 
only accelerating charges emit light'' \cite{Schwartz}, and the construction  
should be put in the context of what was known and understood in 1690.  
Yet, a principle of equivalence guarantees the validity of Huygens' 
results: given two adjacent domains, 1 and 2, and given a wave in
domain 1 approaching domain 2, it is possible to define conditions at
the boundary between 1 and 2 (charges oscillating at the proper
frequency, amplitude and direction) such that the wave excited in
domain 2 by these boundary sources is indistinguishable from the wave
entering from domain 1. Note that the charges in question oscillate as
if responding to the wave in domain 1. The only imperfection in
Huygens’ construction is that it pictured point-emitters, whereas
elementary dipoles are more appropriate, as discussed in Ref.\cite{Miller} and 
in Sec.\ref{SecNumMeth}. 

In 1816 Fresnel postulated that Huygens' secondary wavelets 
interfere with each other \cite{Fresnel1,Fresnel2}. 
At this point, we can look at the interference pattern, and identify 
the next wavefront as the locus of points with the same phase. 

The combination of Huygens' construction with the principle of 
interference is called Huygens-Fresnel principle \cite{BornWolf} or, often, 
simply Huygens' principle, and is an important principle in the theory of 
diffraction. 

Hence, not surprisingly, the principle has been used for 
didactic physics widgets illustrating diffraction around an obstacle, 
diffraction through one or more slits \cite{Physlet}, 
and the diffraction of a Gaussian beam (namely, its finite waist). 
The principle also accounts for refraction, provided wavelets expand with 
different velocities in different media, in inverse proportion with the 
respective indices of refraction. For this reason, Huygens' principle 
has also been applied to Snell's law \cite{PhysletSnell}. 

Recently Huygens' principle was also invoked in the design of metamaterials 
with new beam shaping, steering and focusing capabilities \cite{HuygMetamat}. 

In addition, the Huygens' construction and Huygens-Fresnel's principle 
inspired numerical methods, briefly reviewed in the next two subsections. 
This is also not surprising, on considering that Huygens' original drawings 
\cite{Huygens} basically outlined an iterative graphical solver, vaguely 
reminescent of a modern iterative solver on a computer.

\subsection{Seismic wavefront tracing methods}

Huygens' principle was considered in Ref.\cite{Mex}, but discarded 
due to the lack of physics rigor mentioned above \cite{Schwartz}. 

Wavefronts are ``traced'' or ``tracked'' 
in seismology on the basis of ray-tracing 
results or of geometrical-optics-approximated calculations 
of shortest travel-time to points on a grid \cite{Rawlinson}, which in 
optics is equivalent to Fermat's principle of least time.  
In particular, among grid-based methods, fast marching methods \cite{FMM} 
and some earlier works \cite{QinLuo} are 
close to Huygens' idea, in that they treat the wavefront as a moving interface. 
Ray tracing ignores diffraction by definition, but 
calculations of shortest travel-time interrogate all points in a domain, 
including points around an obstacle. This might look promising for the 
inclusion of diffraction effects. Yet, it should be pointed out that these 
methods typically solve the Wentzel-Kramers-Brillouin (WKB) approximated 
eikonal equation. Therefore, the wave field is only accurate in the 
high-frequency, short-wavelength limit. 

An interesting and elegant method is developed in Refs.\cite{Sava1,Sava2}. 
Again, the starting point is the eikonal equation, but manipulated into the 
expression:
\begin{equation}    \label{eqExpSph}
  [x-x(\gamma,\phi)]^2 + 
  [y-y(\gamma,\phi)]^2 + 
  [z-z(\gamma,\phi)]^2 = r^2(\gamma,\phi), 
\end{equation}
where $\gamma$ and $\phi$ are curvilinear coordinates on the wavefront, 
$x(\gamma,\phi)$, $y(\gamma,\phi)$, $z(\gamma,\phi)$ are the Cartesian 
coordinates of a (source) point on the wavefront, and $x,y,z$ the 
coordinates of another point, which can be pictured as a receiver. 
Eq.\ref{eqExpSph} describes a sphere centered at 
$x(\gamma,\phi),  y(\gamma,\phi), z(\gamma,\phi)$, of radius 
$r(\gamma,\phi)=v(\gamma,\phi)\tau$, where $v$ is a velocity. 
As time $\tau$ progresses, the sphere expands. 
Differentiating Eq.\ref{eqExpSph} eventually leads 
to an explicit finite-difference scheme: the coordinates of a point 
on the new wavefront are functions of the local $r$ and of the coordinates 
of some points on the current wavefront. In particular, stencils of three and 
five points are used respectively in the 2D and 3D problem. 
The analogies with Huygens' construction are evident. However, 
Eq.\ref{eqExpSph} contains no information on the strength of 
individual point-sources. 
Also, each new point is only informed by three or five points in its vicinity, 
whereas diffraction integrals imply that every point on the new wavefront 
depends on every point on the old one. Hence, although the numerical  
technique is highly stable \cite{Sava1,Sava2}, 
small phase-increments might be required in between consecutive wavefronts, 
in order for the restriction to three or five points to be a good 
approximation.

\subsection{Other wavefront tracing and Huygens-related methods}

Transmission Line Matrix Modeling \cite{Hoefer} 
maps the problem of wave propagation in a medium 
into the problem of electrical signal propagation in a 3D array of shunt and 
series nodes. It can be shown that this is equivalent to Huygens' mechanical 
model of light as a perturbation of the Ether, originated by intense heat and  
transmitted by elastic shocks from one particle to the next, in all directions  
\cite{Hoefer}. It is also equivalent to a many-body scattering problem, 
in which each node scatters light emitted by other nodes. It is possible to 
include diffraction  
by proper formulation of the scattered wave, or by adding {\em ad hoc}  
dielectrics to the array mentioned above \cite{Marche}. 

Face offsetting \cite{FaceOffset} 
is a Lagrangian algorithm that treats an interface as a surface-mesh,  
propagates the mesh faces and, from their envelope, reconstructs the next 
surface, and its vertices. The method is based on a generalization of 
Huygens' original construction and envelope idea, but without 
Fresnel's addition of wavelet-interference. 

Diffraction by rough surfaces (thus, ultimately, Huygens-Fresnel principle) 
is the physical explanation for iridescent colors. However, 
their rendering in computer graphics is simply obtained by angle-dependent 
coloration of surfaces in ray tracings \cite{Agu}. 

Recent algorithms \cite{Luo} utilize the Huygens-Kirchhoff integral to 
integrate many locally valid asymptotic Green functions into a globally valid 
asymptotic Green function for the Helmoltz equation in inhomogeneous media. 
However, they are formulated under geometrical optics approximations. 

Additional methods and potential numerical applications of Huygens' 
principle are discussed in Ref.\cite{EuroPhys}. 

Note that ``wavefront tracing'' here refers to reconstructing 
several wavefronts in a propagation problem, whereas 
``wavefront reconstruction'' in large telescopes denotes the inverse 
problem (phase retrieval) of reconstructing, from sensor measurements, 
a single incoming wavefront, 
in order to decide how to make it planar by means of adaptive optics 
\cite{SIAM}. Yet, there might be synergies with the method proposed here, 
one step of which is indeed phase retrieval (Sec.\ref{}).  
Wavefront reconstruction includes diffraction.

\subsection{Present work}

In summary, although several numerical works use Huygens’ construction
of emitters and wavelets, few of them utilize the Huygens-Fresnel
principle in its entirety, including wavelet interference (that is,
diffraction). 

A very simple numerical method is presented in Sec.\ref{SecNumMeth} 
of the present article, that retains more than 3 or 5  
points in the diffraction integrals, 
takes into account their different intensities and utilizes a zero-search 
to localize the points forming the new wavefront. 
Sec.\ref{SecNumExamples} presents numerical examples obtained in a 2D 
isotropic non-uniform medium. 

The original motivation for this work stemmed from the search for wave 
propagation methods in magnetized plasmas, intermediate in physics content  
and computational cost between ray-tracing \cite{Tracy,Batchelor,VolpeRSI} 
and beam-tracing or paraxial Wentzel-Kramers-Brillouin (WKB) 
methods \cite{EPoli, Bertelli} on one hand, and 
full-wave solvers \cite{Brambilla,Jaeger} on the other. In turn, such need 
was prompted by unexpected results from the first full-wave study in the 
electron cyclotron frequency range \cite{Vdovin} -a high-frequency range, 
typically well-modeled by ray tracings. 
However, the method proposed here is not restricted to magnetized plasmas, 
and could be extended 
to generic 3D anisotropic birefringent media, as discussed in 
Sec.\ref{SecFutW}.

\section{Numerical Method} \label{SecNumMeth}

\subsection{Governing equation}
Consider a wave in a medium, of 
wavenumber $k=2\pi/\lambda$, where $\lambda$ is the wavelength 
in the medium. In general, $k$ and $\lambda$ differ from the wavenumber 
and wavelength in vacuum.

The Huygens-Fresnel formula for the complex field amplitude $\bf E$ in a
point $\bf x'$ can be written as follows \cite{Makris}: 
\begin{equation}  \label{EqHuygens}
    {\bf E} ({\bf x'}) = 
    \frac{1}{4\pi} \int_S 
    \left[
      \frac{\cos\theta}{r} - ik(1+\cos\theta)
    \right] 
    \frac{e^{ikr}}{r} {\bf E} ({\bf x}) dS,  
\end{equation}
where $\bf x$ is the generic point on surface $S$, representing the initial 
wavefront. $\theta$ denotes the
angle between the local wavefront-normal and the vector $\bf x'$-{\bf x}, 
of norm $r$. 

Strictly speaking, Eq.\ref{EqHuygens} is only correct for uniform media, 
which, however, can often also be treated analytically.  
Nonetheless, it is not unusual and perhaps more useful and interesting 
to also use Eq.\ref{EqHuygens}  
\cite{Kravtsov, Baues, Monzon, Ghatak, Gitin} 
or other Huygens-based expressions \cite{Rawlinson, Sava1, Sava2, Luo, PBJohns} 
in weakly non-uniform media (in the sense that these media 
can be treated as locally 
uniform on the lengthscale of a single numerical step, but could 
have different refractive properties in the following step, and the wave have a 
different $k$ and $\lambda$ as a result). 

Said otherwise, the formula is still applicable if the steplength is properly 
chosen, so that $\bf x '$ is ``not too far'' from any $\bf x$, in the sense 
that $L = \left( \frac{1}{k} \frac{dk}{dx} \right)^{-1} \gg |\bf x'$-{\bf x}$|$. 
Here $L$ is the lengthscale over which the 
medium (hence, the wavenumber $k$) varies. 
Note that the constraint on $\bf x'$ not being too far depends on $L$. 
Also note that the approximation 
does not necessarily depend on how $L$ 
compares with the wavelength $\lambda$=$2\pi/k$. In other words, even if 
the medium varies significantly over a wavelength $\lambda$, 
Eq.\ref{EqHuygens} can still be used to calculate $\bf E$ in a point $\bf x'$ 
close enough to $S$ (``close enough'' in the sense that the medium is 
practically uniform over that short distance). In fact, any 
ordering of the wavelength $\lambda$ with respect to $L$ and 
$|\bf x'$-{\bf x}$|$ is acceptable: greater than both, smaller than both, 
or intermediate.

In 2D, the surface
integral is replaced by a line integral, and the denominator by $\sqrt{r}$,
because the intensity of a cylindrical wave decays like 1/$r$ 
\cite{Luo,Nonogaki}: 
\begin{equation}  \label{EqHuygensLin}
    {\bf E} ({\bf x'}) = 
    e^{-i\pi/4}
    \int_l 
    \sqrt{ \frac{1+\cos \theta}{2\lambda} }
    \frac{e^{ikr}}{\sqrt{r}} {\bf E} ({\bf x}) dl.
\end{equation}
The integrand is more complicated than expected from Huygens'
simplistic point sources. To begin with, the elementary sources emit
with different intensities in different directions. This was
conjectured by Fresnel to reconcile theory and experiment, and was
rigorously calculated by Stokes \cite{BornWolf}. 
It responds to the intuition that,
given a source and two observers, one with a direct line of view, the
other located around a corner, the former should receive more light.

Another correction is the $\cos \theta /r$ term in Eq.\ref{EqHuygens}. 
This is needed to make
Huygens-Formula valid also in the near-field, and consistent with the
more general and rigorous Kirchhoff integral theorem \cite{Miller}. Its
physical meaning is that the elementary sources should actually be
radiating dipoles.  

Indeed, dipoles are routinely used in modeling
meta-lenses, frequency-selective surfaces and other diffraction-based
devices \cite{Mudar,Capasso,Ken}. 

Note that Eqs.\ref{EqHuygens} and \ref{EqHuygensLin} are {\em not} equations to 
solve. They are rather {\em solutions}, in 3D and 2D, 
of the Helmholtz equation governing 
electromagnetic wave propagation in uniform (and, with some approximation, 
 {\em slowly} non-uniform) media. With special care, they can also describe 
propagation in strongly non-uniform media (see Figs.\ref{FigRefr} and 
\ref{FigSnell} and accompanying discussion). 
 
These solutions are constructed as superpositions of point-to-point solutions, 
propagating from a point-emitter to a point-observer. 
The treatment presented in the present paper permits to efficiently calculate 
integrals \ref{EqHuygens} or \ref{EqHuygensLin} in realistic domains with  
obstacles and non-uniformities. Its outcome are wavefronts reconstructed with 
high precision -strongly sub-wavelength. Yet, these finely reconstructed 
wavefronts can be sought at large distances from each other (large compared 
with the wavelength, but small compared with the lengthscale of inhomogeneity). 

The treatment presented is easily generalized to other problems, 
provided the point-to-point propagator (Green's function) is known, 
and the superposition principle is valid.

\subsection{Wavefront construction by zero localization} \label{}
Given a wavefront $S$, our objective is to compute the integral in 
Eq.\ref{EqHuygens} 
to identify a set of points ${\bf x'}_{ab}$ (with $a$=1,\dots $m$ and 
$b$=1,\dots $n$) where $\bf E$ 
has equal phase. Without losing generality, we can search for
$\Re({\bf E})$=0: if ${\bf E}={\bf A}e^\phi$, then $\Re({\bf E})$=0 implies 
either zero amplitude ($\bf A=0$) or a specific phase ($\cos \phi$=0).  
However, the trivial solution of zero amplitude is easily rejected: 
zero amplitude means no wave, and if there is no wave in $\bf x'$, 
presumably there is no wave in a neighborhood of $\bf x'$. 
By contrast, $\Re({\bf E})$=0 as due to $\cos \phi$=0 can only be true on a 
particular wavefront.

This search for a specific phase ($\phi=\pi/2$ modulo $\pi$) 
is easily generalized 
to a different phase by shifting all sources in the integrand by a 
certain $\delta\phi$. The points found form the new wavefront.  

Equivalently, one
could search for the locus of points where the phase of the total
electric field, $\arctan [\Im({\bf E})/\Re({\bf E})]$ 
equals the phase of interest,
$\phi$. This can also be casted as a zero-search problem.  

The existence of 
multiple solutions is prevented by restricting the search to intervals
of length $\lambda$/2, where $\lambda$ is the wavelength in the medium.
It should be clarified that the {\em length} 
of the search-interval has nothing to do with its {\em positioning}: 
the search-interval can be placed close to the 
wavefront $S$ or far from it, depending on the desired resolution 
(in the sense of inter-wavefront spacing) and on 
the validity of Eq.\ref{EqHuygens} at that distance from $S$ 
(in other words, is it still $|{\bf x}'-{\bf x}|\ll \left[ \frac{1}{k} 
\frac{dk}{dx} ({\bf x}) \right]^{-1}$ for any ${\bf x}'$ on the $\lambda/2$ long 
search-interval and for every $\bf x$ on surface $S$?).
Note that the wavefront-to-wavefront distance depends on the 
positioning of the search-intervals, not on their length.  
An important consequence is that 
acceptable wavefront-to-wavefront distances (thus, 
computational costs) do not depend on 
$\lambda$; they only depend on the inhomogeneity of the medium. 

Depending on where the user places the search-intervals, the 
algorithm will trace the wavefront at 
phase $\phi=\pi/2$, or $3\pi/2$, or $5\pi/2$ etc. from the current one. 
As discussed, this is because sources have 
zero initial phase in Eq.\ref{EqHuygens}. 
Nonetheless, different wavefront-to-wavefront phase-increments (different from 
odd multiples of $\pi/2$) can also be obtained. For this purpose we can 
perturb the phase at the sources 
($e^{ikr} {\bf E} ({\bf x}) \rightarrow e^{ikr + i\phi} {\bf E} ({\bf x}) $) 
while still searching for zeros at the receivers ($\Re [{\bf E}({\bf x}')]=0$).

At least two approaches can now be envisioned for either zero search 
(that is, of $\Re ({\bf E})$ or 
$\arctan [\Im ({\bf E})/\Re({\bf E})]-\phi$). 

A possible approach is to locally solve Maxwell's equations 
in such volume, in presence of point-sources located on the current wavefront.
Computational electromagnetics methods, including but not limited 
to finite differences and finite elements, could 
repetitively solve the ``local'' problem in front of the present wavefront. 
However, in general this approach offers 
no major advantage compared to solving the problem once in the ``global'' 
domain. The approach is only advantageous if the union of the local domains 
is smaller than the global domain.   

More interestingly, and 
more in line with the spirit of the Huygens-Fresnel principle, 
we can let the original point-sources interfere and we can localize iso-phase 
points to construct the new wavefront. In particular, we can look for loci of 
destructive interference (zeros).

\subsection{Varying the phase-step} \label{}
The zero-search is conducted on a $\lambda$/2 long segment. Depending on the
positioning of the search intervals, different zeros could be found,
corresponding to a different wavefront or, which is the same, to a
different phase-advance $\Delta \phi$ (modulo $\pi$) 
with respect to the previous
front. It is necessary to consistently place the search-intervals in
such a way that they bracket zeros all belonging to the same
wavefront. In other words, the search intervals need to ``bracket'' the
unknown wavefront with a precision of $\lambda$/2.  

This provides a simple “knob” to vary the phase-advance from one
wavefront to the next: it is sufficient to consistently move all
search-intervals farther from or closer to the current wavefront to
vary the phase-advance by $n\pi$, where $n$ is an integer.  

For finer adjustments of $\phi$, one can use complex $\bf E$ in the 
integrand of Eq.\ref{EqHuygens} or, equivalently, multiply the integrand 
by $e^{i\phi}$, which is equivalent to phase-shifting the sources.

\subsection{Amplitude calculation and initialization of following iteration}
If the wavefront was constructed as a locus of zeros ($\Re({\bf E})$=0), it
would be unclear what the wave amplitude $\bf E$ is in each point. This
information is needed because the final wavefront of a given 
step is the initial wavefront for the following one, and its points need 
to be properly initialized with the correct amplitude. 

For this, it will be necessary to either compute $\Im({\bf E})$ in the
wavefront points (if $\Re({\bf E})$ is already known), 
or re-compute $\Re({\bf E})$ in those same points, but with
all sources phase-shifted by $\phi \neq n\pi$. 
Note that the wavefront points have
already been identified and do not need to be identified again. 

Also note that this helps discarding ``false zeros'' of the wave-field. 
By this we mean that, in addition to zeros ($\Re({\bf E})\simeq$0) 
of the oscillating, non-negligible wave-field ($|{\bf E}|\neq$0), 
there could be trivial zeros where $\Re({\bf E})\simeq$0 
and $|{\bf E}|\simeq$0. Such zeros are located away from the wave beam, and 
are not relevant to wavefront identification.

\subsection{Algorithm}   \label{SecAlgo}
The algorithm -in 2D to fix the ideas, in Cartesian coordinates $x$ and
$y$, but easily generalized to 3D- can be recapitulated and illustrated
as follows.

\begin{figure}[t]
\begin{center}
           \includegraphics[scale=1.5]{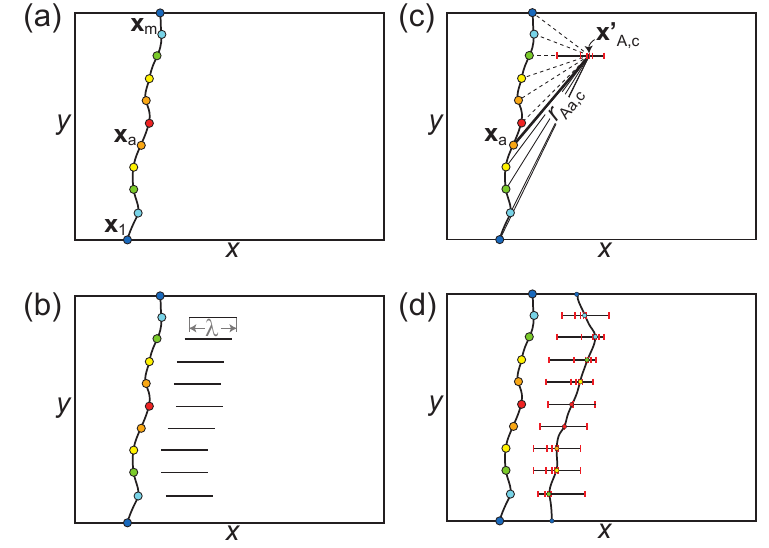}
         \caption{Illustration of (a) step 1, (b) step 2 and (c-d) step 3 of 
           the algorithm described in Sec.\ref{SecAlgo}, for tracing a new 
           wavefront based on the current one.\label{FigSketches}}
\end{center}
\end{figure}

{\bf Step 1}: Discretize the current wavefront as a
1D array of points ${\bf x}_a$ ($a$=1,\dots $m$) 
emitting different intensities ${\bf E}_a$ 
(color-coded in Fig.\ref{FigSketches}a), 
distributed according to the beam-pattern.

The very first array can be at the antenna, phased array or last
mirror and can have complicated shape or phase-pattern. 

{\bf Step 2}: Define $\lambda$/2 long search-intervals for the points 
${\bf x}'_A$ ($A$=1,\dots $m$) that will form the next wavefront.  

For simplicity they can have the same coordinates $y$ as the previous points 
${\bf x}_a$, and only span $x$ (Fig.\ref{FigSketches}b). 
In fact, they can be obtained by displacing all points ${\bf x}_a$  by
the same amounts $\Delta x$ and $\Delta x +\lambda$/2. 
Avoiding to place such intervals too closely ($\Delta x \ll \lambda$) 
to the original wavefront will allow discarding the dipole
correction. 

{\bf Step 3}: For $A$ going from 1 to $m$: 
\begin{enumerate}[label=\alph*)]
    \item 
    Consider the search-interval for ${\bf x}'_A$.  
    \item 
    Evaluate the field ${\bf E}_{Ac}$ in a point
    ${\bf x}'_{Ac}$ on that interval, where $c$ denotes an iteration index in
    the zero-search, e.g. by bisection \cite{BookBisect}. 
    Note that ${\bf E}_{Ac}$ (of fixed $A$) is obtained by interference of 
    all point-sources ($a$=1,\dots $m$) in Fig.\ref{FigSketches}c, 
    each one of amplitude ${\bf E}_a$, at distance
    $r_{Aa,c}$ from the target and inclination $\theta_{Aa,c}$: 
    \begin{equation}  \label{EqHuygensLinDiscr}
        {\bf E}_{Ac} = 
        e^{-i\pi/4}
        \sum_{a=1}^{m} 
        \sqrt{ \frac{1+\cos \theta_{Aa,c}}{2\lambda} }
        \frac{e^{ikr_{Aa,c}}}{\sqrt{r_{Aa,c}}} {\bf E}_a \Delta l_a
    \end{equation}
    This is a discretization of Eq.\ref{EqHuygensLin}. 
    In its present form, the number of field calculations in $m$ observers 
    due to $m$ emitters scales like $m^2$. However, 
    unequally spaced Fast Fourier Transform methods \cite{Rokhlin} 
    could make it scale like $m \log m$. 
    Additionally, Eq.\ref{EqHuygensLinDiscr} can be restricted to values of $a$
    (typically in the neighborhood of $A$) yielding non-negligible
    contributions (dashed lines in Fig.\ref{FigSketches}c).  
    
    \item 
    Repeat step b) for
    various ${\bf x}'_{Ac}$ ``suggested'' by the zero-finder in its convergence
    to the zero. Repeat until $|\Re({\bf E}_{Ac})|<\epsilon$, 
    where $\epsilon$ is a prescribed tolerance.  
    
    \item
    Set point ${\bf x}'_A$ equal to last ${\bf x}'_{Ac}$.  
    
    \item 
    Set field-amplitude in ${\bf x}'_A$ 
    equal to last $\Im ({\bf E}_{Ac})$ (Fig.\ref{FigSketches}d).
\end{enumerate}

{\bf Step 4}: Repeat steps 1-3 until absorption, or until exiting
from computational domain, or until other criterion is met. Use the
final points and amplitudes of step 3 as initial points and amplitudes
for step 1.

The 2D algorithm is easily generalized to 3D by adding a subscript $_b$ 
and summing and looping over it. Eq.\ref{EqHuygensLinDiscr} 
needs to be replaced by:
    \begin{equation}  \label{EqHuygensDiscr}
        {\bf E}_{ABc} = - 
        \sum_{b=1}^{n} 
        \sum_{a=1}^{m}
        \frac{ik}{4\pi}
        (1+\cos \theta_{AB,ab,c})
        \frac{e^{ikr_{AB,ab,c}}}{r_{AB,ab,c}}
        {\bf E}_{ab} \Delta S_{ab}.
    \end{equation}

\section{Numerical examples} \label{SecNumExamples}
\begin{figure}[t]
\begin{center}
           \includegraphics[scale=1.5]{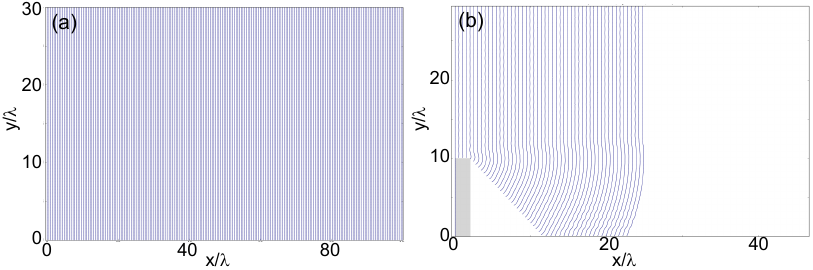}
         \caption{(a) Simple test of tracing planar wavefronts in uniform 
           medium, in the absence of obstacles. 
           (b) Initially planar wavefronts diffract 
           around obstacle as expected. \label{FigObstacle}}
\end{center}
\end{figure}

The first test consisted of aiming a plane wave in the 
$x$ direction, with periodic boundary conditions in $y$, effectively 
simulating an infinitely wide planar wavefront. 
This very simple test worked as expected: the wavefront propagated  
in the $x$ direction and remained planar (Fig.\ref{FigObstacle}a).
This is in obvious agreement with the trivial analytical solution. 

Next, an obstacle was placed at the bottom left of the computational box 
(Fig.\ref{FigObstacle}b). 
No periodic boundary conditions were used here, but additional emitters 
(``ghost cells'') were added at large $y$, outside of the computational box 
of interest. These acted as 
point-sources, which were taken into account in localizing new zeros and 
constructing new wavefronts, as well as calculating the amplitudes in 
wavefront points, for proper initialization of each step. 
Fig.\ref{FigObstacle} clearly exhibits diffraction around the obstacle, 
as expected. 
At shallow angles past the obstacle, as expected, the wave field is weak 
and its zeros are not shown. 

\begin{figure}[t]
\begin{center}
   \includegraphics[scale=1.5]{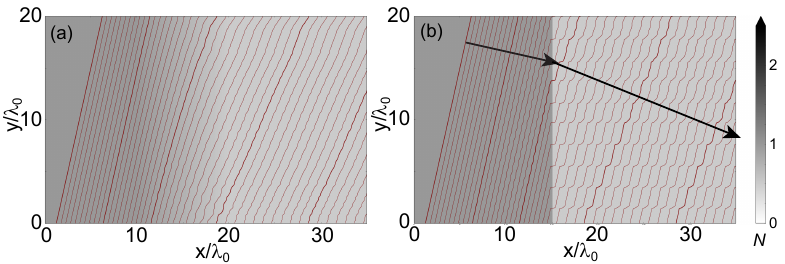}
   \caption{Wavefronts refracting in a medium of 
 (a) refractive index $N$ varying over ten wavelengths or (b) discontinuous.
   \label{FigRefr}}
\end{center}
\end{figure}

The next problem modeled was refraction in a non-uniform medium whose 
index of refraction varies slowly with space (ten wavelengths, in the 
case of Fig.\ref{FigRefr}). The wavefront is planar, oblique. It extends  
outside of the computational domain by means of ghost cells on its 
top and bottom. As expected, the wavefront changes its direction of 
propagation according to the index of refraction, in agreement with the 
continous limit of Snell's law. The wavelength also contracts or 
expands accordingly, as it can be recognized from the wavefront spacing. 

\begin{figure}[t]
\begin{center}
   \includegraphics[scale=0.16]{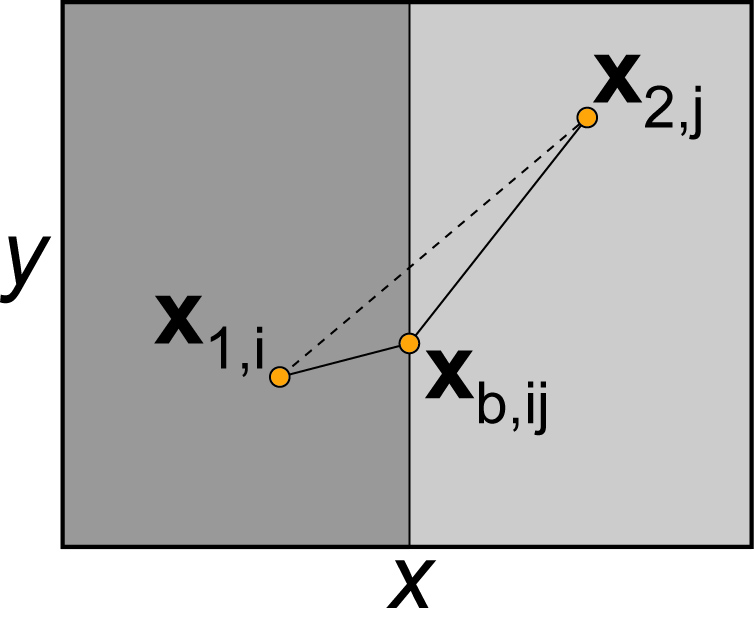}
   \caption{Illustration of why, in presence of discontinuities,  
     a single Green's function cannot propagate the information 
     from an emitter in medium 1 to a receiver in medium 2, and an intermediate 
     point is needed at the boundary, determined by Snell's law. 
   \label{FigSnell}}
\end{center}
\end{figure}

Discontinuities in the refractive index require special care. 
The reason is that the generic wavefront crossing a discontinuity 
has some points in medium 1, 
and some in medium 2. If $\bf E$ is being evaluated in a point in medium 2, 
the contributions from points in medium 1 will travel initially 
at speed $c/N_1$ and then, once reached medium 2, at speed $c/N_2$, 
where $N_1$ and $N_2$ are the refractive indices in the two media. 
Due to the principle of least time, the information does not travel on a 
straight line, but on two consecutive segments of different 
inclinations, obeying Snell's law. Therefore, instead of a single Green's 
function relating emitter $i$ with observation point $j$, 
\begin{equation}
    \frac{e^{i{\bf k}\cdot ({\bf x}_j-{\bf x}_i)}}{|{\bf x}_j-{\bf x}_i|}, 
\end{equation}
we should introduce an intermediate point ${\bf x}_{b,ij}$ at the boundary, 
and use two propagators, from ${\bf x}_{1,i}$ in medium 1   
to ${\bf x}_{b,ij}$, and then from ${\bf x}_{b,ij}$ to ${\bf x}_{2,j}$ in medium 2:
\begin{equation}
    \frac{e^{i{\bf k}_2\cdot ({\bf x}_{2,j}-{\bf x}_{b,ij})}}{|{\bf x}_{2,j}-{\bf x}_{b,ij}|}
    \frac{e^{i{\bf k}_1\cdot ({\bf x}_{b,ij}-{\bf x}_{1,i})}}{|{\bf x}_{b,ij}-{\bf x}_{1,i}|}
\end{equation}
For a given pair of ${\bf x}_{1,i}$ and ${\bf x}_{2,j}$, 
${\bf x}_{b,ij}$ is uniquely determined by Snell's law (Fig.\ref{FigSnell}). 

The wavefront bends as expected from Snell's law (Fig.\ref{FigRefr}b), in 
agreement with the analytical result. 
Corrugations similar to Fig.\ref{FigRefr}a, but stronger, 
are ascribed to a lack of ghost cells, and will be the subject of future work. 
The construction of search-intervals will also be improved, by better 
bracketing of the zeros of interest.

\begin{figure}[t]
\begin{center}
   \includegraphics{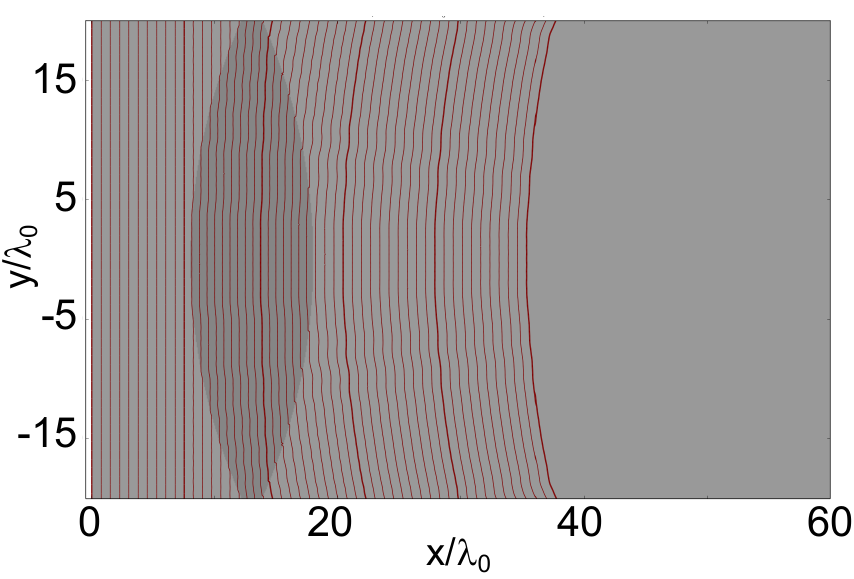}
   \caption{Wavefronts focused by a convergent lens of refractive index 
     $N$=1.2. 
   \label{FigLens}}
\end{center}
\end{figure}

Related to refraction by a single discontinuity is refraction by two 
consecutive ones. Wavefronts experience this when crossing a lens, 
and are observed to deform as expected, for example in the case 
of a convergent lens of refractive index $N$=1.2 (Fig.\ref{FigLens}). 

Propagation through a single lens can be easily generalized to 
propagation through a random medium consisting of variously shaped 
``blobs'' or grains immersed in a background material of different 
refractive index.

\begin{figure}[t]
\begin{center}
   \includegraphics[scale=1.5]{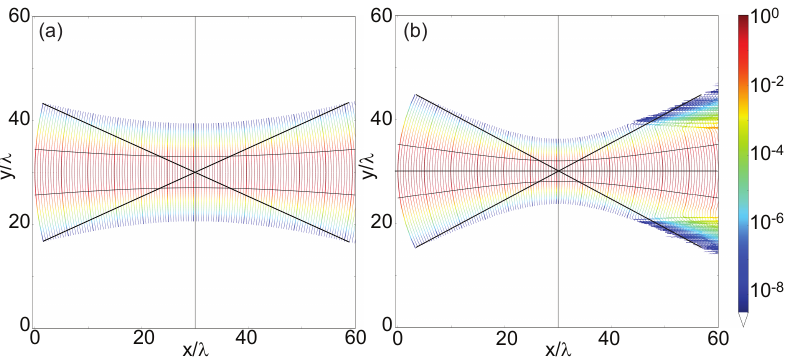}
   \caption{Wavefronts for (a) moderately and (b) strongly focused 
   Gaussian beam, colored according to intensity normalized to on-axis value. 
   One in every ten wavefronts is bold, as a guide for eyes. Also shown are the 
   exact solutions for the contours at 1/$e^2$ of central intensity (black 
   curves on red background).
   \label{FigGauss}}
\end{center}
\end{figure}

Finally, a Gaussian beam was simulated. 
Point-emitters were initialized on an initial curved wavefront, with 
intensities distributed according to a Gaussian profile. 
Excellent results were obtained for a moderately focused Gaussian beam 
(Fig.\ref{FigGauss}a), 
in good agreement with Gaussian optics theory (curves in Fig.\ref{FigGauss}a) 
\cite{Goldsmith}. 
The geometrical optics solution is also shown and, as expected, 
is a good approximation away from the waist.

Fig.\ref{FigGauss} does not show wave intensities smaller than 
$10^{-8}$, when normalized to intensities on the axis of the Gaussian beam. 
This corresponds to fractional electric fields of $10^{-4}$. Similar 
criteria were adopted in truncating the very edge of the wavefront 
diffracting around the obstacle in Fig.\ref{FigObstacle}. 

At even smaller fields, the zero-search can fail 
and misplace portions of the wavefront. 
Although this error originates in typically uninteresting low-$E$ regions, 
it should be noted that each wavefront is determined by the previous one. 
Hence, imprecisely traced portions of a wavefront might cause 
portions of ``later'' wavefronts to also be misplaced, or distorted.  
The ultimate result is that 
these errors can eventually affect finite-$E$ regions.  
This is typically not the case for moderately focused beams 
(Fig.\ref{FigGauss}a), but can 
be the case for the strongly focused ones (Fig.\ref{FigGauss}b), which 
will be the subject of future work. 
The issue could be solved by an improved, adaptive definition 
of the interval for zero-search by the bisection method. Additionally, 
here the zero search was performed for simplicity in the $x$ direction 
(see search-intervals in Fig.\ref{FigSnell}b). However, more customised 
search-intervals (orthogonal to the wavefront, or, better, parallel to 
group velocity) would aim at regions of high $E$ and avoid low 
$E$ and associated issues.

\section{Discussion} \label{SecDisc}
By virtue of Fresnel's principle, the method naturally includes diffraction. 
This is an advantage over ray tracings and {\em truncated} 
paraxial expansions used in beam-tracings. 

Another advantage of the method is that it only requires sub-wavelength 
resolution {\em along} the wavefront. {\em Across} the wavefront, however, the 
search for the next wavefront only requires a handful of points: 
if we assume the next wavefront to be one wavelength $\lambda$ away from the 
current one, then only $\log_2 100$ iterations, -that is, 6 or 7- are needed   
to localize a point on the next wavefront with $\lambda/100$ accuracy. 
This is an important advantage over methods needing 100 points or elements 
per wavelength. 
Moreover, the next wavefront can be searched at more than 
a wavelength from the previous one, resulting in a further speed-up. 
This is possible because complete 
diffraction formulas are used here, instead of near-field approximations. 

Going in more details, let us estimate the number of point-to-point 
calculations $N_{ops}$ needed to trace $N_{WT}$ wavefronts discretized by 
$N_E$ points. Here by point-to-point we mean the contribution of a specific
point emitter to a specific observation point (the argument 
of the sums in Eq.\ref{EqHuygensDiscr}). 
Localizing each point requires calculating
the total field according to  Eq.\ref{EqHuygensDiscr} 
in  $N_{bis}$ “trial-points”. These are the points needed by the
bisection method to converge to a zero. In turn, each total field is
calculated as the sum over a total of $N_E$ emitters, in the worst case 
(i.e., on the net of truncations).
Hence, a total of $N_{ops}=N_E^2 N_{WT} N_{bis}$  point-to-point
calculations is needed. 

Wavefront-spacing is
not constrained by the wavelength, but needs to be much smaller than
the non-uniformity length-scale normally to the wavefront, $L_\perp$. 
Nevertheless, it is reasonable to assume that wavefront-spacing is a
fraction $f$ of the wavelength. Therefore, $N_{WT}=N_{\lambda \perp} /f$, where 
$N_{\lambda \perp}$ is the number of wavelengths in the direction orthogonal to 
the wavefront.  
In that case, reaching the orthogonal precision of a full-wave solver 
deploying $N_{ppw}$ points per unit wavelength imposes 
$N_{bis}=\log_2 (f N_{ppw})$.  
In conclusion, $N_{ops}=N_E^2 N_{\lambda \perp} \log_2 (f N_{ppw})/f$.

As mentioned in Sec.\ref{SecAlgo}, unequally spaced Fast Fourier Transform 
methods \cite{Rokhlin} could further speed up the method by replacing 
$N_E^2 \rightarrow N_E \log N_E$.

\section{Future work} \label{SecFutW}
In the original Huygens-Fresnel integral, $k$ was a fixed scalar, but
{\em non-uniformity} is easily taken into account by letting $k=k(x)$, as 
done in Sec.\ref{SecNumMeth}.

Also, the term  in Huygens’ integrand assumed isotropy: $k$ is a scalar,
meaning that wavelets expand in all directions ${\bf x'}-{\bf x}$ 
with the same wavelength and same phase velocity $\omega/k$.  

In {\em anisotropic} media, however, global
wavefronts as well as local wavelets expand at different velocities in
different directions. Spherical wavelets are thus replaced by
ellipsoids, elongated or compressed in the direction of
anisotropy. Such ellipsoids do not preserve their aspect ratio,
because some axes expand faster than others. It is easy to account for
this by letting $k=k({\bf x}, {\bf x}'-{\bf x})$. 
The wavenumber now depends on the line-of-sight ${\bf x'}-{\bf x}$  
connecting the emitter with the observer (for
example it depends on whether such direction is parallel,
perpendicular or oblique to the direction of anisotropy). $k$ is provided by 
the dispersion relation for the wave of interest, in the polarization
(mode) of interest, and in the direction ${\bf x'}-{\bf x}$ of interest. 
An earlier
treatment of anisotropy in terms of TE and TM modes \cite{Bergstein} 
arrived to the same ellipsoidal wavelet representation. 

Finally, further generalizing ${\bf k}={\bf k}({\bf x}, {\bf x'-x}, mode)$ 
to be a complex, mode-dependent vector, not necessarily parallel to 
${\bf x'-x}$, accounts respectively for {\em absorption}, 
{\em birefringence} and {\em decoupling between phase and group velocity}. 
The phase velocity is parallel to $\bf k$, the group velocity describes 
energy propagation from an array of points ${\bf x}$ to an array of points 
$\bf x'$. Incidentally, phase and group velocity are decoupled also in ray 
tracings, allowing for non-parallel increments of ray position $d\bf r$ and 
of wavevector, $d\bf k$. 

The underlying theory for  ${\bf k}={\bf k}({\bf x}, {\bf x'-x}, mode)$  
(dispersion relation) 
is very well-established, but was never inserted in Huygens’ formulas.

Further improvements can be realized by 
modifying the integrand in Eqs.\ref{EqHuygens}-\ref{EqHuygensLin} 
by way of special functions:
\begin{itemize}
  \item Ref.\cite{Sava1} proposed to treat {\em caustics} 
    by strategically ignoring some 
    points of the old wavefront. In Ref.\cite{Kravtsov} it 
    was proposed to address caustics 
    by replacing with Airy functions the field radiated by point-sources. 
  \item Bessel functions, on the other hand, can model the diffraction of 
    {\em evanescent waves} in the near field \cite{Makris}. 
  \item Airy and parabolic cylinder functions can replace oscillating 
    fields in some evanescent {\em mode-conversion} regions \cite{VolpeOLA}. 
\end{itemize}

\section*{Summary and Conclusions}
We explored the applicability of the Huygens-Fresnel principle 
to a new method for wave propagation in non-uniform media. 

Very briefly, the wavefront is discretized in an array of
point-emitters of wavelets. A zero-search identifies observation
points where the wavelets interfere destructively. Those zeros of
field form an iso-phase surface, that is easily generalized to other
phases. Then the process is repeated.

Some very simple cases were successfully modeled in 2D, including diffraction 
around an obstacle, refraction in slowly-varying and discontinuous media, 
Gaussian beams.   

Minor issues of phase-jumps and corrugations were encountered, and 
fixed respectively by improved selection of zero-search intervals, 
and by adding ``ghost cells'' outside of the computational domain. 

Future work was outlined and discussed, regarding extensions 
to anisotropic, birefringent, absorbing media such as magnetized plasmas.

\section*{Acknowledgements}
The first author gratefully acknowledges fruitful discussions with 
R.~Bilato, M.~Brambilla, O.~Maj and E.~Poli, all at IPP Garching, Germany.



\bibliographystyle{elsarticle-num} 

\begin{thebibliography}{10}
\expandafter\ifx\csname url\endcsname\relax
  \def\url#1{\texttt{#1}}\fi
\expandafter\ifx\csname urlprefix\endcsname\relax\def\urlprefix{URL }\fi
\expandafter\ifx\csname href\endcsname\relax
  \def\href#1#2{#2} \def\path#1{#1}\fi

\bibitem{Huygens}
C.~Huygens, Trait\'{e} de la Lumiere, Leyden, 1690.

\bibitem{BornWolf}
M.~Born, E.~Wolf, Principles of Optics, Cambridge University Press, 2006.

\bibitem{Fresnel1}
A.~Fresnel, Ann.\ Chem.\ Phys. 1 (1816) 239.

\bibitem{Fresnel2}
A.~Fresnel, Mem.\ Acad. 5 (1826) 339.

\bibitem{Baker}
B.~Baker, E.~Copson, The Mathematical Theory of Huygens' Principle, AMS Chelsea
  Publishing, 1987.

\bibitem{Schwartz}
M.~Schwartz, Principles of Electrodynamics, Dover, Mineola, NY, 1987.

\bibitem{Miller}
D.~Miller, Optics Lett. 16 (1991) 1370.

\bibitem{Physlet}
P.~Falstad, \url{http://www.falstad.com/ripple/} (2014).

\bibitem{PhysletSnell}
W.~Fendt, \url{http://www.walter-fendt.de/ph14e/huygenspr.htm} (2010).

\bibitem{HuygMetamat}
C.~Pfeiffer, A.~Grbic, Phys.\ Rev.\ Lett. 110 (2013) 197401.

\bibitem{Mex}
C.~Lomnitz, Y.~Meas, Geophys.\ Res. Lett. 31 (2004) L13613.

\bibitem{Rawlinson}
N.~Rawlinson, J.~Hauser, M.~Sambridge, Advances in Geophys.\ 49 (2008) 203.

\bibitem{FMM}
M.~de~Kool, N.~Rawlinson, M.~Sambridge, Geophys.\ J.\ Int. 167 (2006) 253.

\bibitem{QinLuo}
F.~Qin, Y.~Luo, K.~Olsen, W.~Cai, G.~Schuster, Geophysics 57 (1991) 478.

\bibitem{Sava1}
P.~Sava, S.~Formel, SEP Report 95 (1997) 101.

\bibitem{Sava2}
P.~Sava, S.~Formel, Geophysics 66 (2001) 883.

\bibitem{Hoefer}
W.~Hoefer, Proc. IEEE 79 (1991) 1459.

\bibitem{Marche}
L.~Pierantoni, A.~Massaro, T.~Rozzi, A tlm node for the diffraction by
  3d-dielectric corners based on the simultaneous transverse resonance method,
  in: IEEE MTT-S International Microwave Symposium digest. IEEE MTT-S
  International Microwave Symposium, 2005, pp. 1077--1080.

\bibitem{FaceOffset}
X.~Jiao, J. Comput. Phys. 220 (2006) 612.

\bibitem{Agu}
E.~Agu, F.~Hill, A simple method for ray tracing diffraction, in:
  Proc.~ICCSA'03 Proceedings of the 2003 international conference on
  Computational science and its applications: PartIII, 2003, pp. 336--345.

\bibitem{Luo}
S.~Luo, J.~Qian, R.~Burridge, J. Comput. Phys. 270 (2014) 378.

\bibitem{EuroPhys}
P.~Enders, Eur. J. Phys. 17 (1996) 226.

\bibitem{SIAM}
D.~Luke, J.~Burke, R.~Lyon, SIAM Rev. 44 (2002) 169.

\bibitem{Tracy}
E.~Tracy, A.~Brizard, A.~Kaufman, A.~Richardson, Ray Tracing and Beyond: Phase
  Space Methods in Plasma Wave Theory, Cambridge Univ. Press, Cambridge, 2014.

\bibitem{Batchelor}
D.~Batchelor, R.~Goldfinger, H.~Weitzner, IEEE Trans.\ Plasma Sci. PS-8 (1980)
  78.

\bibitem{VolpeRSI}
F.~Volpe, H.~Laqua, Rev.\ Sci.\ Instrum. 74 (2003) 1409.

\bibitem{EPoli}
E.~Poli, A.~Peeters, G.~Pereverzev, Comp.\ Phys.\ Comm. 136 (2001) 90.

\bibitem{Bertelli}
N.~Bertelli, O.~Maj, E.~Poli, R.~Harvey, J.~Wright, P.~Bonoli, C.~Phillips,
  A.~Smirnov, E.~Valeo, J.~Wilson, Phys. Plasmas 19 (2012) 08510.

\bibitem{Brambilla}
M.~Brambilla, Plasma Phys. Control. Fusion 41 (1999) 1.

\bibitem{Jaeger}
E.~Jaeger, L.~Berry, E.~D’Azevedo, et~al., Phys.\ Plasmas 15 (2008) 072513.

\bibitem{Vdovin}
V.~Vdovin, Fusion Science \& Technology 59 (2011) 690.

\bibitem{Makris}
K.~Makris, D.~Psaltis, Optics Communications 284 (2011) 1686.

\bibitem{Kravtsov}
Y.~Kravtsov, Z.~Feizulin, Radiophysics and Quantum Electronics 12 (1969) 706.

\bibitem{Baues}
P.~Baues, Opto-Electronics 1 (1969) 37.

\bibitem{Monzon}
J.~Monzon, IEEE Trans. Microwave Theory Tech. 41 (1993) 1995.

\bibitem{Ghatak}
A.~Ghatak, Optics, McGraw-Hill, 2009.

\bibitem{Gitin}
A.~Gitin, Applied Optics 52 (2013) 7419.

\bibitem{PBJohns}
P.~Johns, IEEE Trans. Microwave Theory Tech. MTT-22 (1974) 209.

\bibitem{Nonogaki}
S.~Nonogaki, Japanese Journ. Appl. Phys. 28 (1989) 786.

\bibitem{Mudar}
M.~Al-Joumayly, N.~Behdad, IEEE Trans. Antennas Prop. 58 (2010) 4033.

\bibitem{Capasso}
J.~Tetienne, R.~Blanchard, N.~Yu, et~al., New Journal of Physics 13 (2011)
  053057.

\bibitem{Ken}
K.~Hammond, S.~Massidda, W.~Capecchi, F.~Volpe, J. Infrared Milli. Terahz Waves
  34 (2013) 437.

\bibitem{BookBisect}
R.~Burden, J.~Faires, A.~Burden, Numerical Analysis, Cengage Learning, 2015.

\bibitem{Rokhlin}
A.~Dutt, V.~Rokhlin, SIAM J. Sci. Comput. 14 (1993) 1368.

\bibitem{Goldsmith}
P.~Goldsmith, Quasioptical Systems, Wiley, 1998.

\bibitem{Bergstein}
L.~Bergstein, T.~Zachos, Journ. Opt. Soc. America 56 (1966) 931.

\bibitem{VolpeOLA}
F.~Volpe, Phys.\ Lett.\ A 374 (2010) 1736.

\end{thebibliography}





\end{document}